\documentclass[aps,prl,twocolumn,superscriptaddress,english]{revtex4}
\usepackage{amssymb}
\usepackage{graphicx}
\usepackage{amsmath}
\usepackage{amsthm}
\usepackage{amsfonts}
\usepackage[T1]{fontenc}
\usepackage[latin9]{inputenc}
\usepackage{array}
\usepackage{multirow}
\usepackage{color}
\usepackage{esint}
\usepackage{bm}
\usepackage{color}
\usepackage{bbm}
\usepackage{hyperref}
\usepackage{babel}
\usepackage{titlesec}
\usepackage{changes}

\begin{document}

\global\long\def\id{\mathbbm{1}}
\global\long\def\ui{\mathbbm{i}}
\global\long\def\ud{\mathrm{d}}

\title{One dimensional quasiperiodic mosaic lattice with exact mobility edges}

\author{Yucheng Wang}
\thanks{These authors contribute equally to this work.}
\affiliation{Shenzhen Institute for Quantum Science and Engineering, and Department of Physics,
Southern University of Science and Technology, Shenzhen 518055, China}
\affiliation{International Center for Quantum Materials, School of Physics, Peking University, Beijing 100871, China}
\affiliation{Collaborative Innovation Center of Quantum Matter, Beijing 100871, China}
\author{Xu Xia}
\thanks{These authors contribute equally to this work.}
\affiliation{Chern Institute of Mathematics and LPMC, Nankai University, Tianjin 300071, China}
\author{Long Zhang}
\affiliation{International Center for Quantum Materials, School of Physics, Peking University, Beijing 100871, China}
\affiliation{Collaborative Innovation Center of Quantum Matter, Beijing 100871, China}
\author{Hepeng Yao}
\affiliation{CPHT, CNRS, Institut Polytechnique de Paris, Route de Saclay 91128 Palaiseau, France}
\author{Shu Chen}
\affiliation{Beijing National Laboratory for Condensed Matter Physics, Institute of Physics,
Chinese Academy of Sciences, Beijing 100190, China}
\affiliation{School of Physical Sciences, University of Chinese Academy of Sciences, Beijing,
100049, China}
\affiliation{Yangtze River Delta Physics Research Center, Liyang, Jiangsu 213300, China}
\author{Jiangong You}
\email{jyou@nankai.edu.cn}
\affiliation{Chern Institute of Mathematics and LPMC, Nankai University, Tianjin 300071, China}
\author{Qi Zhou}
\email{qizhou@nankai.edu.cn}
\affiliation{Chern Institute of Mathematics and LPMC, Nankai University, Tianjin 300071, China}
\author{Xiong-Jun Liu}
\email{xiongjunliu@pku.edu.cn}
\affiliation{International Center for Quantum Materials, School of Physics, Peking University, Beijing 100871, China}
\affiliation{Collaborative Innovation Center of Quantum Matter, Beijing 100871, China}
\affiliation{CAS Center for Excellence in Topological Quantum Computation, University of Chinese Academy of Sciences, Beijing 100190, China}
\affiliation{Shenzhen Institute for Quantum Science and Engineering, and Department of Physics,
Southern University of Science and Technology, Shenzhen 518055, China}

\begin{abstract}
The mobility edges (MEs) in energy which separate extended and localized states are a central concept in understanding the localization physics. In one-dimensional (1D) quasiperiodic systems, while MEs may exist for certain cases, the analytic results which allow for an exact understanding are rare. Here we uncover a class of exactly solvable 1D models with MEs in the spectra, where quasiperiodic on-site potentials are inlaid in the lattice with equally spaced sites. The analytical solutions provide the exact results not only for the MEs, but also for the localization and extended features of all states in the spectra, as derived through computing the Lyapunov exponents from Avila's global theory, and also numerically verified by calculating the fractal dimension. We further propose a novel scheme with experimental feasibility to realize our model based on an optical Raman lattice, 
which paves the way for experimental exploration of the predicted exact ME physics.
\end{abstract}
\maketitle

{\em Introduction.--}Anderson localization (AL) is a fundamental and extensively studied quantum phenomenon, in which the disorder induces exponentially localized electronic wave-functions, and results in the absence of diffusion~\cite{Anderson1958}. For the one and two dimensions, the states in the disordered systems are all localized~\cite{Anderson1979}. For a three-dimensional (3D) system, beyond the critical disorder strength, a mobility edge (ME) which marks a critical energy $E_c$ separating extended states from localized ones may be resulted and can lead to novel fundamental physics~\cite{Evers2008}. For instance, varying the disorder strength or particle number density may shift the position of ME across Fermi energy, and induce the metal-insulator transition. Moreover, in a system with ME only the particles of a finite energy window can flow. This can enable a strong thermoelectric response~\cite{Whitney2014,Goold2020,Kaoru2017}, which is widely used in thermoelectric devices. Nevertheless, it is hard to introduce microscopic models to understand the physics of the ME in 3D systems~\cite{Kulkarni2017}, so it is highly important to develop lower dimensional models with MEs, especially with exact MEs which allows for analytical studies.

When the random disorder is replaced by quasiperiodic potential, the system may host localized and delocalized states even in the low dimension regime.
In particular, the extended-AL transitions and MEs have been predicted in 1D quasiperiodic systems~\cite{Xie1988,Hashimoto1992n,Boers2007,Biddle2009,Biddle,Lellouch2014,Li2017,Yao2019,Ganeshan2015,Xu2020,Chen2020,Yong2020}. The simplest nontrivial example with 1D quasiperiodic potential is the Aubry-Andr\'{e}-Harper (AAH) model~\cite{AA}, which shows a phase transition from a completely extended phase to a completely localized phase with increasing the strength of the quasiperiodic potential.
The AAH model exhibits a self duality at the transition point for the transformation between
lattice and momentum spaces. Thus no ME exists for the standard AAH model. However, by introducing a long-range hopping term~\cite{Biddle,Santos2019,Saha2019}, or breaking the self duality of the AAH Hamiltonian, e.g. superposing another quasiperiodic optical lattice~\cite{Li2017,Yao2019,Sun2020} or introducing the spin-orbit coupling~\cite{Zhou2013,Kohmoto2008}, one can obtain MEs in the energy spectra of the system. In very few cases~\cite{Biddle,Ganeshan2015} the self duality may be recovered on certain analytically determined energy, across which the extended-localization transition occurs, rendering the ME in the spectra, while the whole model is not exactly solvable. That is, the extended and localized states in the spectra cannot be analytically obtained to rigorously illustrate how the transition between them occurs. 
In consequence, to introduce and develop more generic models with ME, which can be exactly solved beyond the dual transformation, is highly significant to further explore the rich ME physics.
Moreover, it is not clear if a single system can have multiple MEs, and is important to know what determines the number of the MEs. Addressing these issues with exactly solvable models is critical to gain exact understanding of the extended-localization transition and to advance the in-depth studies of fundamental ME physics, e.g. to possibly eliminate the theoretical dispute that whether the many-body MEs exist~\cite{Roeck2016,Gao2019}.

The quasiperiodic systems can be easily realized in
experiments in ultracold atomic gases trapped by two optical lattices
with incommensurate wavelengths~\cite{Roati2008}. This configuration forms the basis of observing the AL, many body localization, Bose glass ~\cite{Roati2008,Fallani2007,Bloch1,Bordia,Bloch2,Modugno2014,Bloch3,WangYC2019}, and very recently the MEs~\cite{Bloch4,An2018,BlochME5,An2020}. Experimental realization of MEs with analytic functional form can help understanding the ME physics quantitatively and better investigate the effect of novel interacting effects on the MEs~\cite{An2020}. 

In this letter, we propose a class of analytically solvable 1D models in quasiperiodic mosaic lattice, which host multiple MEs with the self-duality breaking.
These models are beyond the conventional ones in which only the MEs but not all the states of the spectra can be precisely determined with dual transformation, and can be exactly solved by applying the profound A. Avila's global theory~\cite{A4}, one of his Fields Medal work, to condensed matter physics.
This theory, beyond the dual transformation, gives an efficient way to calculate the
Lyapunov exponent (LE) of all states. We then obtain analytically not only the exact MEs, which can be multiple here, but also the localization and extended features of all the states in the spectra. 
We further propose a novel scheme with experimental feasibility to realize and detect the exact MEs based on ultracold atoms.

\begin{figure}[t]
\hspace*{-0.5cm}
\includegraphics[width=0.52\textwidth]{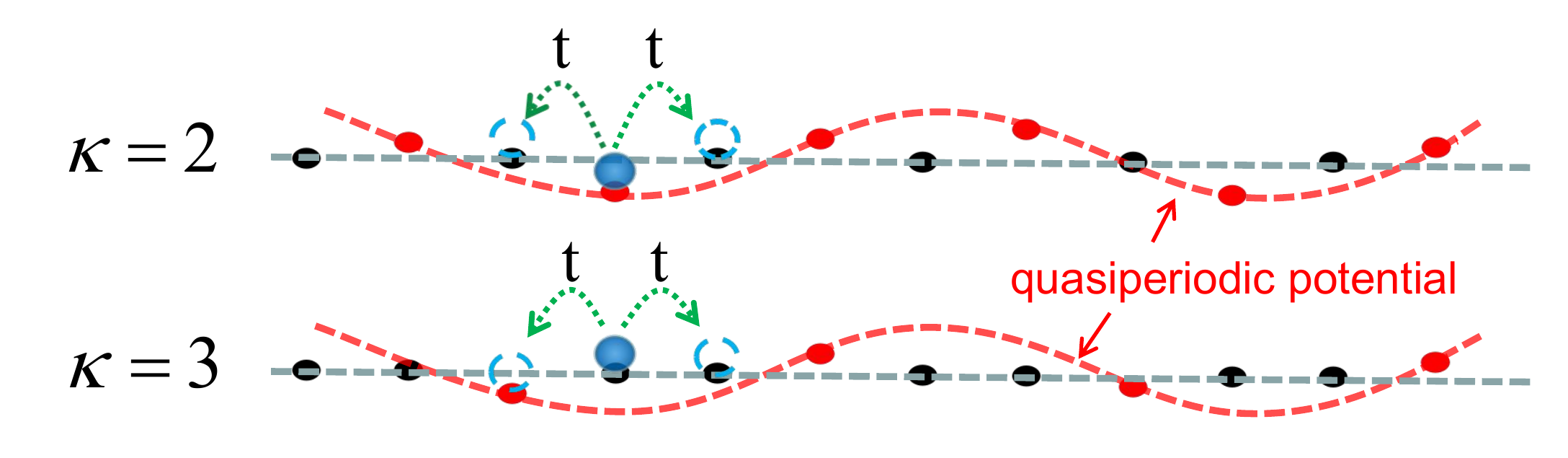}
\caption{\label{01}
The 1D quasiperiodic mosaic model with $\kappa=2$ and $\kappa=3$. The red and black spheres denote the lattice sites whose potentials are quasiperiodic and zero respectively, as shown by the corresponding red and black dashed lines. The blue sphere denotes a particle, and the nearest-neighbor hopping strength is $t$.}
\end{figure}

{\em Model.---}We consider a class of quasiperiodic mosaic models, which can be described by
\begin{eqnarray}\label{ham-1}
H &=& t\sum_{ j}(c^{\dagger}_{j}c_{j+1}+H.c.)+2\sum_j\lambda_jn_{j},
\end{eqnarray}
with
\begin{equation}
\lambda_j=
\begin{cases}
\lambda\cos(2\pi(\omega j+\theta)),\ \  j=m\kappa,  \\
0, \ \textrm{otherwise},
\end{cases}
\end{equation}
where $c_{j}$ is the annihilation operator at site $j$, $n_{j}=c^{\dagger}_{j}c_{j}$ is the local number operator, $t, \lambda, \theta$ denote the nearest-neighbor hopping coefficient, the quasiperiodic potential amplitude, and the phase offset, respectively, $\omega$ is an irrational number, and $\kappa$ is an integer.
We set the hopping strength $t=1$ for convenience. Since the quasiperiodic potential periodically occurs with interval $\kappa$, we can introduce a quasi-cell with the nearest $\kappa$ lattice sites. If the quasi-cell number is taken as $N$, i.e., $m=1, 2,\cdots, N$, the system size will be $L=\kappa N$. The quasiperiodic mosaic model with $\kappa=2$ and $\kappa=3$ are pictorially shown in Fig.~\ref{01}, and other cases are similar.

It is obvious that this model reduces to the AAH model when $\kappa=1$.
If $\kappa\neq 1$, the duality symmetry of these models is broken, which motivate us to show the existence of MEs. In this letter, we prove that these models with $\kappa\neq 1$ do have energy dependent MEs, which are given by the following expression,
\begin{eqnarray}\label{ME1}
|\lambda a_{\kappa}|=1, \ {\mbox{for}} \ E=E_c,
\end{eqnarray}
with
\begin{eqnarray}\label{ak}
a_{\kappa}=
       \frac{1}{\sqrt{E^2-4}}\left( (\frac{E+ \sqrt{E^2-4}}{2})^{\kappa}- (\frac{E- \sqrt{E^2-4}}{2})^{\kappa} \right)
\end{eqnarray}
In addition, all the localized and extended states can be exactly studied. This is our central result which we prove by computing the LE exactly. Before showing the analytic derivatives,
we display the numerical evidence for the $\kappa=2$ and $\kappa=3$ cases, which benefit a visual understanding of this condition (Eq.~\eqref{ME1}) representing it as a ME.
Without loss of generality, we set $\theta=0$ and $\omega=\frac{\sqrt{5}-1}{2}$, which can be approached by using the Fibonacci numbers $F_{n}$~\cite{Kohmoto1983,Wang2016n,WangYC2020}: $\omega=\lim_{n \rightarrow \infty}\frac{F_{n-1}}{F_{n}}$, where $F_{n}$ is  defined recursively by  $F_{n+1}=F_{n-1}+F_{n}$, with $F_0=F_1=1$. We take the system size $L=F_{n}$ and the rational approximation $\omega=F_{n-1}/F_{n}$ to ensure a periodic boundary condition when numerically diagonalizing the tight binding model defined in Eq.~\eqref{ham-1}.

{\em The $\kappa=2$ and $\kappa=3$ cases.---}For the minimal nontrivial case with $\kappa=2$, the two MEs read~\cite{noteak}
\begin{equation}\label{k2}
  E_c =\pm \frac{1}{\lambda}.
\end{equation}
For the $\kappa=3$ case, the four MEs are given by $E_c=\pm \sqrt{1\pm 1/\lambda}$.
The numerical results are obtained from the inverse participation ratio (IPR) ${\rm IPR}(m) = \sum_{j=1}^L |\psi_{m,j}|^4$~\cite{Evers2008}, where $\psi_m$ is the $m$-th eigenstate. To characterize the ME, we investigate the fractal dimension of the wave function, which is given by $\Gamma=-\lim_{L\rightarrow\infty}\frac{\ln({\rm IPR})}{\ln L}$.
It is known that $\Gamma\to 1$ for extended states and $\Gamma\to 0$ for localized states. We plot energy eigenvalues and the fractal dimension $\Gamma$ of the corresponding eigenstates as a function of potential strength $\lambda$ in Fig.~\ref{02}. The dashed lines in the
figure represent the MEs for $\kappa=2$ and $\kappa=3$, respectively. As expected from the analytical results, $\Gamma$ approximately changes from zero to one
when the energies across the dashed lines. Further, for any $\kappa$, one can obtain $2(\kappa-1)$ MEs well described by Eq.~\eqref{ME1} and Eq.~\eqref{ak}.
\begin{figure}[t]
\centering
\hspace*{-0.3cm}
\includegraphics[width=0.5\textwidth]{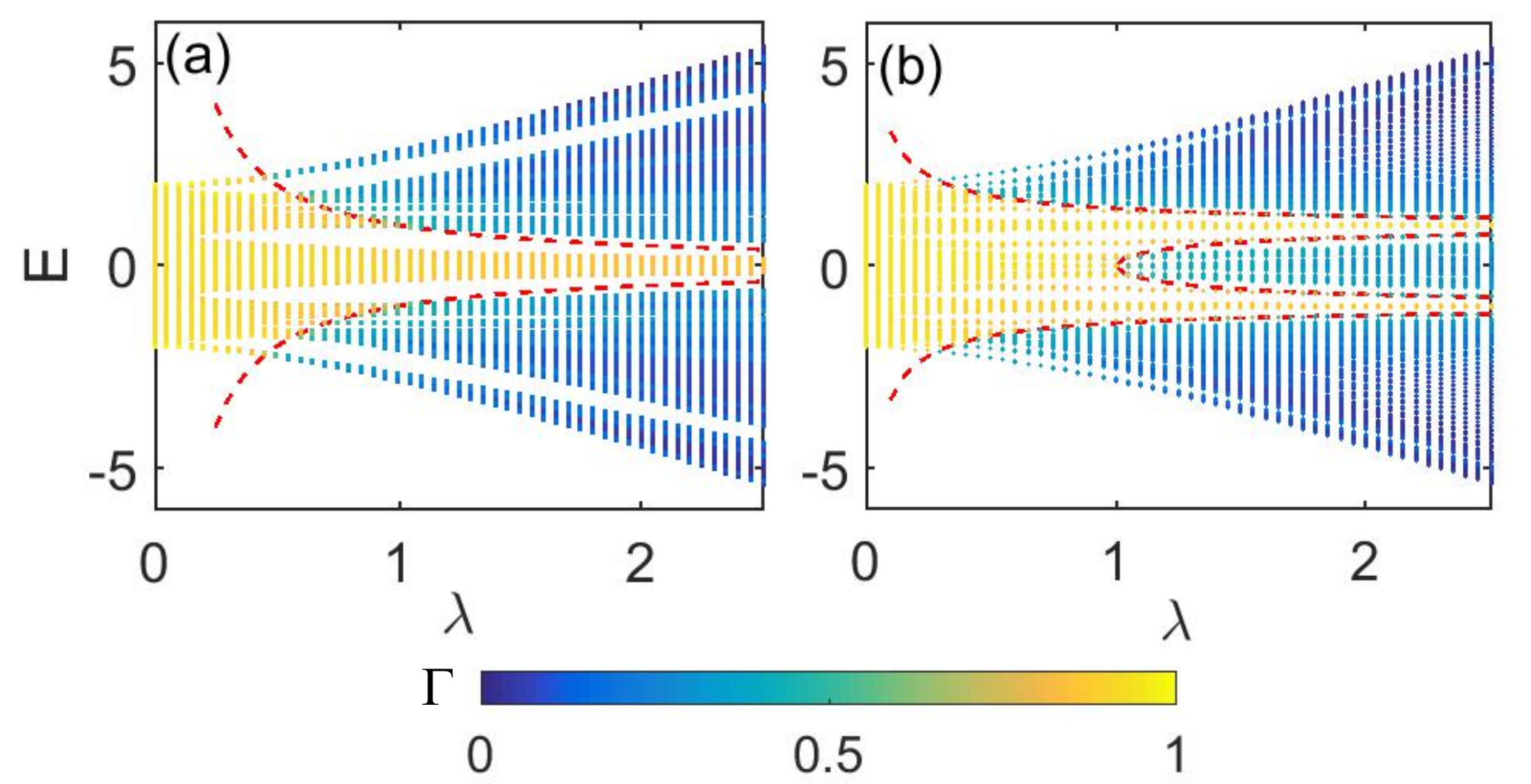}
\caption{\label{02}
 Fractal dimension $\Gamma$ of different eigenstates as a function of the corresponding eigenvalues and quasiperiodic potential strength $\lambda$ for (a) $\kappa=2$ with size $L=F_{14}=610$ and (b) $\kappa=3$ with size $L=F_{15}=987$. The red dashed lines represent the MEs given in Eq.~\eqref{ME1}.}
\end{figure}

The localization starts from the edges of the spectrum, as the coupling constant $\lambda$ is increased, then we have MEs, and for $\kappa=2$ MEs moves towards the center of the spectrum. This behavior is similar to MEs in the 3D disordered systems. However, the present model has a new fundamental feature that in the arbitrarily strong quasiperiodic potential regime, the MEs always take place, i.e, the extended states always exist. This is in sharp contrast to models with random disorder and to other quasiperiodic models, where all the states are localized when the disorder is large enough. In addition, we see that the critical strength of quasiperiodic potential in extended-localization transition of the ground state is smaller than that in the standard AAH model. 
This is because for the mosaic lattice the particle tends to stay at the site with the smallest potential and the potential difference strongly impedes the nearest-neighbor hopping.

\begin{figure}[t]
\centering
\includegraphics[width=0.5\textwidth]{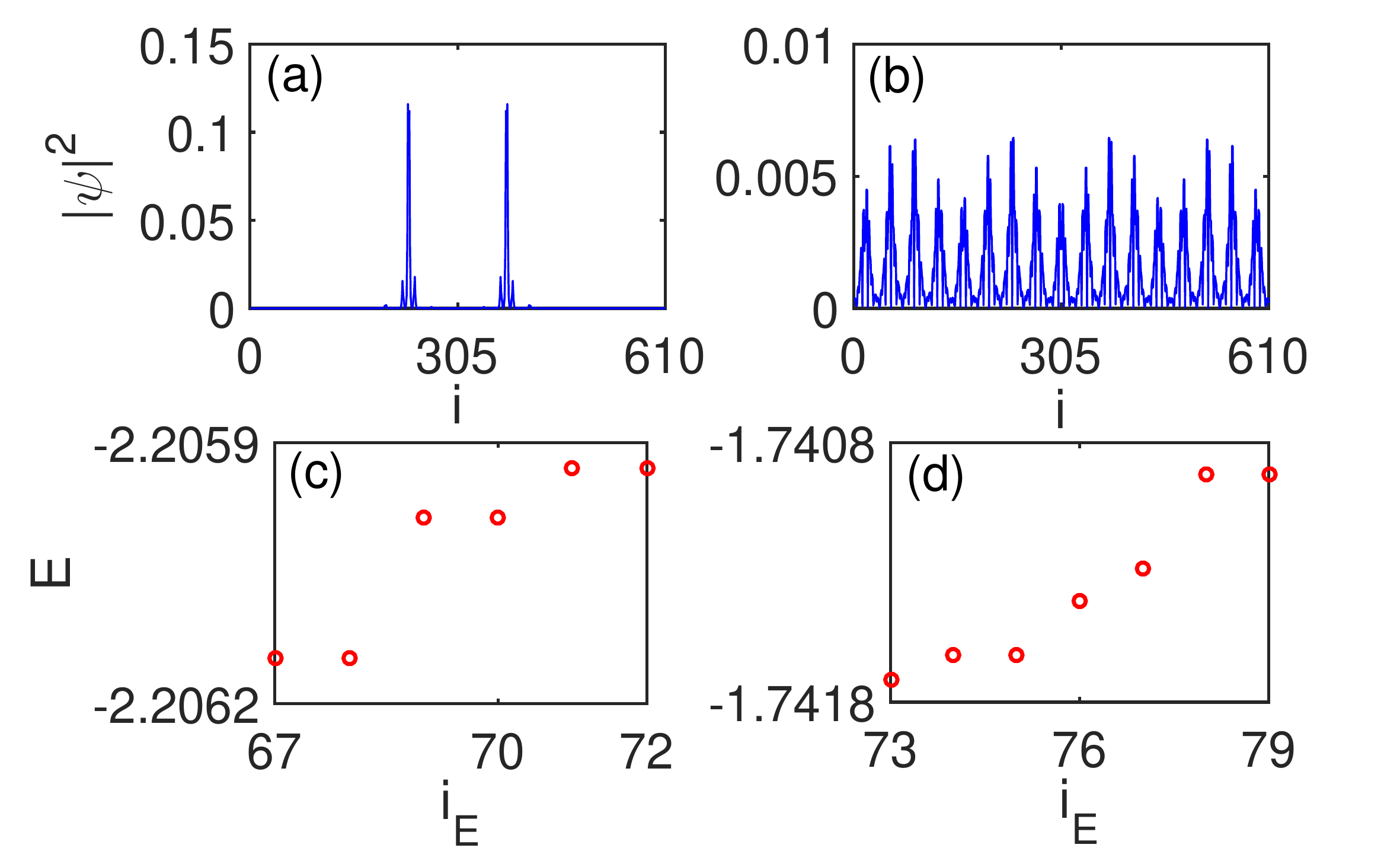}
\caption{\label{03}
Spatial distributions of two eigenstates correspond to (a) $E=-2.205(9)$ and (b) $E=-1.741(7)$, which respectively correspond to the nearest-neighbor eigenvalue below and above the ME of the system. Eigen-energies versus the corresponding index (c) from $67$ to $72$ and (d) from $73$ to $79$, which are respectively below and above the ME ($E_c=-2$), here the eigen-energies in ascending order. Here we fix $\kappa=2$, $\lambda=0.5$ and $L=610$.}
\end{figure}

The ME can be further confirmed by the spatial distributions of the wave
functions, as shown in Fig.~\ref{03} (a) and (b). The wave functions for $\kappa=2$ are localized and extended when their eigenvalues satisfy $|E|>\frac{1}{\lambda}$ and $|E|<\frac{1}{\lambda}$, respectively. It is interesting
that two localization peaks are typically obtained for the localized states [see e.g. Fig.~\ref{03} (a)]. This is due to the existence of two-fold degeneracy of energy levels~\cite{notedeg} which are spatially separated from each other, as shown in Fig.~\ref{03} (c) and (d). We have verified that most of the energy levels are two-fold degenerate for any $\kappa$ greater than $1$. This phenomenon is related to the parent two-fold degeneracy for the $k$ and $-k$ states in the lattice model when there is no quasiperiodic potential. The interesting thing is that while the presence of the inlaid quasiperiodic potential breaks the lattice translational symmetry and the quasimomentum is no longer a good quantum number, the two-fold degeneracy is inherited in the most of the localized states.


{\em Rigorous mathematical proof.---}Now we provide the analytical derivation for the MEs by computing the LE. Denote by $T_n(\theta)$ the transfer matrix of the Schr\"{o}dinger operator~\cite{A4}, then LE can computed as $$\gamma_{\epsilon}(E)=\lim_{n\rightarrow \infty}\frac{1}{n} \int \ln  \|T_n(\theta +i \epsilon)\| d\theta,$$ where $\|A\|$ denotes the norm of the matrix $A$.  The complexification of the phase is important for us, since our computation relies on A.Avila's global theory of one-frequency analytical $SL(2,\mathbb{R})$ cocycle~\cite{A4}.  First note that the transfer matrix can be written as
\begin{equation*}
T_{\kappa}(\theta)= \left(
\begin{array}{cc}
  E-2\lambda \cos 2\pi  (\theta+\kappa \omega) & -1 \\
  1 & 0 \\
\end{array}
\right)
\left(
\begin{array}{cc}
  E& -1 \\
  1 & 0 \\
\end{array}
\right)^{\kappa-1},
\end{equation*}
where \begin{equation*}
\left(
\begin{array}{cc}
  E& -1 \\
  1 & 0 \\
\end{array}
\right)^{\kappa-1} =\left(
\begin{array}{cc}
  a_{\kappa}& -a_{\kappa-1} \\
 a_{\kappa-1} & -a_{\kappa-2} \\
\end{array}
\right)
\end{equation*}
and  $a_{\kappa}$ is defined in \eqref{ak}. Let us then complexify the phase, and let $\epsilon$ goes to infinity,  direct computation yields
\begin{equation}
T_\kappa(\theta+i\epsilon)=e^{2\pi\epsilon}e^{i2\pi(\theta+\kappa \omega)}\left(
\begin{array}{cc}
 -\lambda a_{\kappa} &   \lambda a_{\kappa-1} \\
0 & 0 \\
\end{array}
\right) + o(1).
\notag
\end{equation} Thus we have $\kappa \gamma_{\epsilon}(E)=2\pi\epsilon+\log|{\lambda a_{\kappa}}| +o(1).$
Avila's global theory~\cite{A4,SM} shows that as a function of $\epsilon,$  $\kappa\gamma_{\epsilon}(E)$ is a convex, piecewise linear function, and their slopes are integers multiply $2\pi$. This implies that $\kappa\gamma_{\epsilon}(E)= \max\{\ln|\lambda a_{\kappa}|+2\pi \epsilon, \kappa\gamma_{0}(E)\}.$  Moreover, by Avila's global theory, if the energy does not belong to the spectrum, if and only if $\gamma_{0}(E)>0$, and $\gamma_{\epsilon}(E)$  is an affine function in a neiborghood of $\epsilon=0$.
Consequently, if the energy $E$ lies in the spectrum, we have
$\kappa \gamma_{0}(E)= \max\{\ln| \lambda a_{\kappa}|, 0\}.$  When $|\lambda a_{\kappa}|>1$, $\gamma_0(E)=\frac{\ln| \lambda a_{\kappa}|}{\kappa}$, the state with the energy $E$ is localized has the localization length
\begin{eqnarray}\label{localization}
\xi(E)=\frac{1}{\gamma_0}=\frac{\kappa}{\ln| \lambda a_{\kappa}|},
\end{eqnarray}
which is also verified by numerical results~\cite{SM}.
When $|\lambda a_{\kappa}|<1$, the localization length $\xi\rightarrow\infty$, and the corresponding state is delocalized. Thus
the MEs are determined by $|\lambda a_{\kappa}|=1$ (i.e., Eq.~\eqref{ME1}). In fact, we can further show that the operator has purely absolutely continuous energy spectrum (extended states) for $|\lambda a_{\kappa}|<1$, while it has pure point spectrum for $|\lambda a_{\kappa}|>1$ (localized states)~\cite{paper2020}. This proof also shows the analytic results for the extended and localization features of all the states.


{\em Experimental realization.---} We propose the scheme of realization based on ultracold atoms. We show that the realization of the quasiperiodic mosaic model with $\kappa=2$ is precisely mapped to the realization of a 1D lattice model with spin-$1/2$ atoms, whose Hamiltonian reads
\begin{align}\label{Hlattice}
\begin{split}
&H=\frac{k_x^2}{2m}\otimes\id+{\cal V}_p(x)\sigma_z+M_0\sigma_x+{\cal V}_{s}(x)|\!\downarrow\rangle\langle\downarrow\!|,\\
&{\cal V}_p=\frac{V_p}{2}\cos(2k_px+\phi_p),\,
{\cal V}_{s}=\frac{V_s}{2}\cos(2k_sx+\phi_s),
\end{split}
\end{align}
where $\sigma_{x,y,z}$ are Pauli matrices, ${\cal V}_p(x)$ is a deep spin-dependent primary lattice with spin-conserved hopping being negligible, $M_0$-term couples spin-up and spin-down states, and ${\cal V}_s(x)$ is a secondary incommensurate potential only for spin-down atoms.
One finds that the tight-binding model of $H$ renders the quasiperiodic mosaic model with $\kappa=2$ by mapping the spin-up (spin-down) lattice sites of the former to the odd (even) sites of the latter, the spin-flip coupling $M_0$-term to the hopping $t$-term, and the potential ${\cal V}_s(x)$ to the incommensurate one applied only on odd sites. This basic idea can in principle be generalized to realize quasiperiodic mosaic models of larger $\kappa$ with higher spin systems.

\begin{figure}
\includegraphics[width=0.48\textwidth]{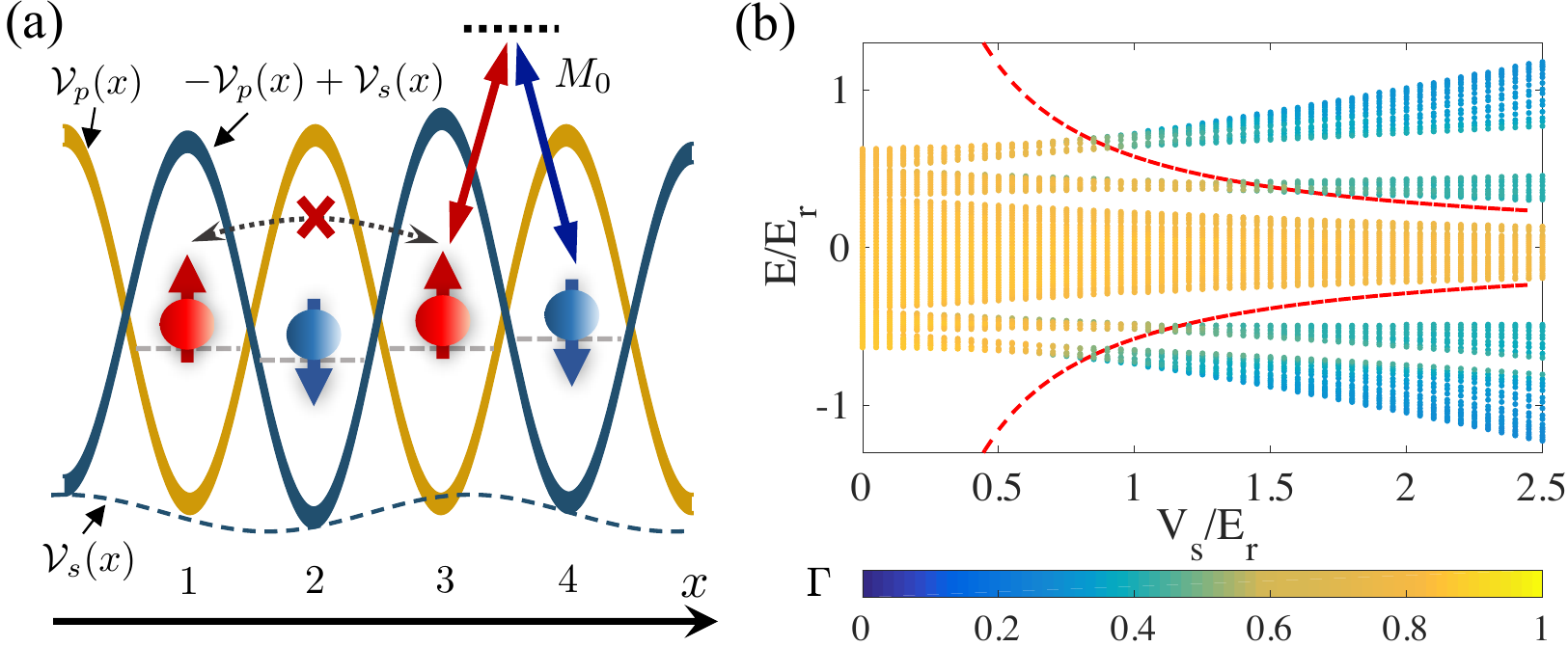}
\caption{Realization of the quasiperiodic model with $\kappa=2$ in cold atoms.
(a) Realization scheme.  The spin-dependent primary lattice ${\cal V}_p(x)$ [$-{\cal V}_p(x)$] locates spin-up (-down) atoms at odd (even) sites,
with an incommensurate potential ${\cal V}_s(x)$ being applied only to the spin-down atoms. The primary lattice is deep enough
such that the spin-conserved hopping can be ignored. A Raman coupling $M_0$ is then used to induce
the spin-flipped hopping, which plays the role of nearest-neighbor tunneling.
(b) Fractal dimension $\Gamma$ of the lowest-band eigenstates of the lattice model as a function of the lattice depth $V_s$.
The eigenvalues $E$ have been shifted a constant value such that
the center of the band is zero for $V_s=0$.
Here we set $V_p=10E_r$, $M_0=1.5E_r$, and $k_s/k_p=\frac{\sqrt{5}-1}{2}$, with $E_r\equiv k^2_p/(2m)$.
The red dashed curves represent the analytical MEs $E_c=\pm t^2/\lambda$, with $t\simeq0.353E_r$ and $\lambda\simeq0.215V_s$.
}\label{fig4}
\end{figure}

The above Hamiltonian can be realized for utracold atoms based on optical Raman lattice [see Fig.~\ref{fig4}(a)]~\cite{WangBZ2018,Song2018,Sun2018}, as briefed below, and the details for the realization are put in Supplementary Material~\cite{SM}.
To facilitate the description, 
we transform the Hamiltonian $H$ with the spin rotation $\sigma_x\to\sigma_z$ and $\sigma_z\to-\sigma_x$.
The primary lattice then reads $-{\cal V}_p(x)\sigma_x$, which induces spin-flip transition in the new bases, and can be generated by two-photon Raman process driven by two laser beams $\bf E_{1,2}$ in the form $\propto{\bf E}^*_1{\bf E}_2\sim\cos(2k_px)$ (see Supplementary Material~\cite{SM}). 
The incommensurate lattice can be similarly obtained by a combination of two potentials $\frac{{\cal V}_{s}}{2}\sigma_x$ and  $-\frac{{\cal V}_{s}}{2}\id$ with ${\cal V}_{s}(x)=\frac{V_s}{2}\sin(2k_sx)$, of which the former is a two-photon Raman coupling potential induced by another two standing-wave beams ${\bf E}_{3,4}$ in the form $\propto{\bf E}^*_3{\bf E}_4$, with $({\bf E}_{3},{\bf E}_{4})\sim(\cos(k_s x),\sin(k_s x))$, while the latter is a standard spin-independent lattice. 
Finally, the $M_0$-term is directly given by the two-photon detuning ($\delta$) of the Raman coupling processes, taking the form $(\delta/2)\sigma_z$. After performing the inverse spin-rotation transformation on these terms, we reach the Hamiltonian~\eqref{Hlattice}. 
More details can be found in Supplementary Material~\cite{SM}, where $^{40}$K atoms is employed to illustrate the realization.

Finally we estimate the parameter regimes for the realization. In experiment, one should set a large $V_p$ compared with $(M_0,V_s)$, such that the spin-conserved hopping $t_p$ (mimicking the next-nearest-neighbor hopping) is negligible.
For example, when $V_p=10E_r$ and $M_0=1.5E_r$ with $E_r\equiv k^2_p/(2m)$, we have $t\simeq18.3t_p$~\cite{SM}.
Thus, regardless of the atom spin and taking into account only $s$-bands, this lattice Hamiltonian (\ref{Hlattice}) indeed leads to the tight-binding model described by Eq.~\eqref{ham-1} with $\kappa=2$.
To further verify our realization scheme,
we calculate the fractal dimension $\Gamma$ of the lowest-band eigenstates of the Hamiltontian (\ref{Hlattice}), and show the results as a function of the lattice depth $V_s$ in Fig.~\ref{fig4}(b).
It can be seen that the distributions of localized and extended states are very similar to the results in Fig.~\ref{02} (a).
We then check the analytical expressions for MEs: $E_c=\pm t^2/\lambda$, where the nearest-neighbor tunneling $t$
and the quasiperiodic potential strength $\lambda\propto V_s$ can be derived based on
$s$-band Wannier functions in the tight-binding limit~\cite{SM}.
We plot the results as red dashed curves in Fig.~\ref{fig4}(b), and find them in good agreement with the fractal dimension calculations.
In experiment, one can determine the MEs by observing the time evolution of an initial charge-density wave state~\cite{Bloch4},
detecting the interference pattern~\cite{Fallani2007}, or characterizing the correlation length~\cite{Modugno2014,Yao2020}.


{\em Conclusion.---}We have proposed a class of exactly solvable 1D mosaic models to realize MEs in energy spectra, where quasiperiodic on-site potentials are inlaid in the lattice with equally spaced sites, and proposed the experimental realization.
By calculating the Lyapunov exponents, we have analytically demonstrated the existence of MEs and obtained their expressions, which are in excellent agreement with the numerical studies. For the integer inlay parameter $\kappa>1$ of our proposed models, one obtains $2(\kappa-1)$ MEs, which are symmetrically distributed in energy spectra and always exist even in the strong quasiperiodic potential regime.
Our work uncovers a variety of new lattice models which host multiple exact MEs and opens a new avenue to analytically explore novel ME physics with experimental feasibility.

We thank Laurent Sanchez-Palencia for helpful comments. Y. Wang, L. Zhang and X.-J. Liu are supported by National Nature Science Foundation of China (11825401, 11761161003, and 11921005), the National Key R\&D Program of China (2016YFA0301604), Guangdong Innovative and Entrepreneurial Research Team Program (No.2016ZT06D348), the Science, Technology and Innovation Commission of Shenzhen Municipality (KYTDPT20181011104202253), and the Strategic Priority Research Program of Chinese Academy of Science (Grant No. XDB28000000). S. Chen was supported by the NSFC (Grant No. 11974413) and the NKRDP of China (Grants No. 2016YFA0300600
and No. 2016YFA0302104). H. Yao acknowledges the support from the Paris region DIM-SIRTEQ. X. Xia is supported by NanKai Zhide Foundation. J. You was partially supported by NSFC grant (11871286) and Nankai Zhide Foundation.  Q. Zhou was partially supported by support by NSFC grant (11671192,11771077) and Nankai Zhide Foundation.

%

\begin{thebibliography}{36}
\bibitem{noteak} We note that in the limit $|E|=2$, we shall find $|a_{\kappa}(\pm 2)|=\kappa$.
\bibitem{notedeg} The Fig.~\ref{03} (a) plots the wave function of one of the two degenerate states. We note that by properly chossing linear combinations of the two degenerate levels, one can construct two orthogonal states with each having only one peak.
\end{thebibliography}

\global\long\def\id{\mathbbm{1}}
\global\long\def\ui{\mathbbm{i}}
\global\long\def\ud{\mathrm{d}}

\setcounter{equation}{0} \setcounter{figure}{0}
\setcounter{table}{0} 
\renewcommand{\theparagraph}{\bf}
\renewcommand{\thefigure}{S\arabic{figure}}
\renewcommand{\theequation}{S\arabic{equation}}

\onecolumngrid
\flushbottom
\newpage
\section*{\large Supplementary Material:\\One dimensional quasiperiodic mosaic lattice with exact mobility edges}
In the Supplementary Materials, we first perform the multifractal analysis. Then, we give
the mathematical basis of computing the Lyapunov exponent and numerical results of the localization length. Finally, we give some details of experimental realization.

\section{I. Multifractal analysis}
To further strengthen the calculated mobility edges (MEs) with $\kappa=2$, we further study the scaling behavior of eigenstates by performing a multifracal analysis~\cite{Kohmoto2008,Kohmoto1989,Jun2016}.
For a normalized eigenstate, the scaling index $\alpha_j$ is defined by $n_j\sim L^{-\alpha_j}$. If this state is extended, for all the lattice sites $j$, we have $\alpha_j\rightarrow 1$ when $L\rightarrow\infty$. If this state is localized, there exist the non-vanishing probabilities only on a finite number of
sites even when $L\rightarrow\infty$, i.e., $\alpha_j\rightarrow0$ for these sites but $\alpha_k\rightarrow\infty$ for remaining sites $k$ with $n_k=0$. Therefore, to identify the extended and localized eigenstates, one can simply examine the minimal value of $\alpha$, which takes $\alpha_{min}=1$ (extended) or $\alpha_{min}=0$ (localized) in the thermodynamic limit.
Fig. \ref{02} (a) and (b) display the $\alpha_{min}$ of two typical eigenstates corresponding to the bottom and center of the spectrum, respectively. We see that the $\alpha_{min}$ tend to $1$ for $\lambda=0.4$, indicating that both of them are extended. In contrast, $\alpha_{min}$ tend to $0$ and $1$ for the ground state and the center state of this system, respectively, with $\lambda=0.5$ and $\lambda=10$, signifying the existence of MEs.
Fig. \ref{02} (c) shows the $\alpha_{min}$ of all eigenstates with size $L=F_{17}=2584$.
There exist dramatic changes of $\alpha_{min}$ at MEs given by Eq.5 ($E_c=\pm \frac{1}{\lambda}$) in the main text with increasing eigenvalues $E$, suggesting
that the predicted MEs well separate localized states from extended states.

\begin{figure}[h]
\hspace*{-0.1cm}
\includegraphics[width=0.9\textwidth]{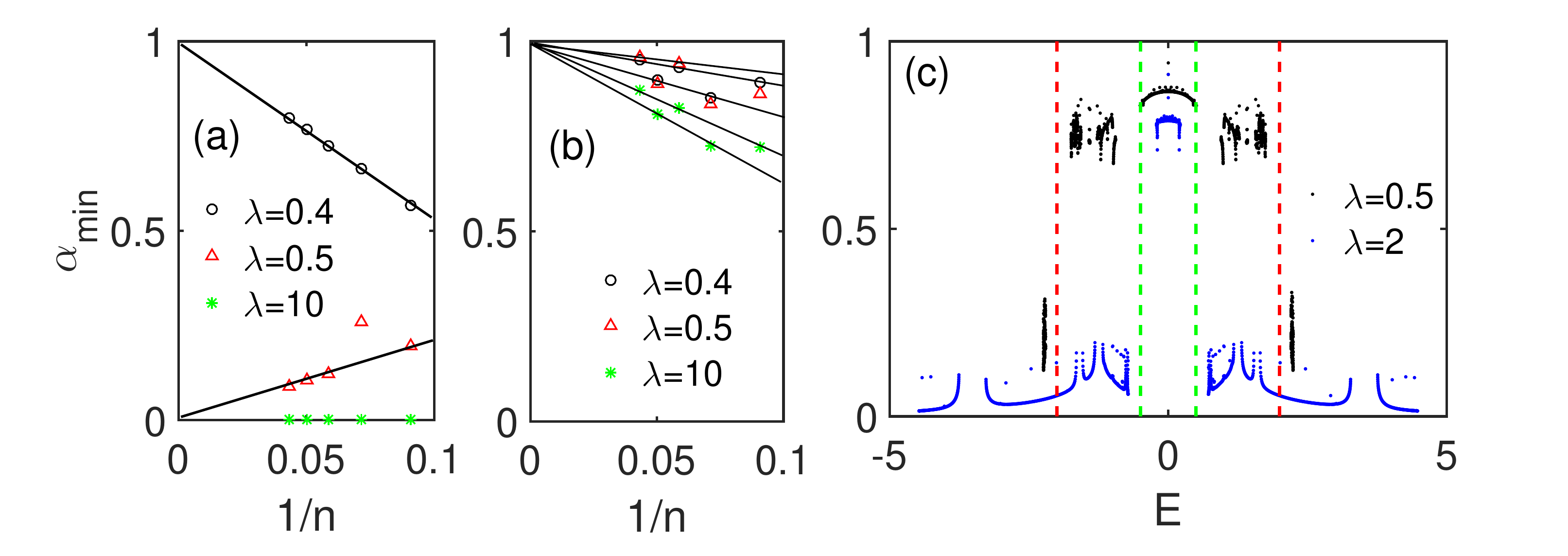}
\caption{\label{02}
 $\alpha_{min}$ as a function of $1/n$ for (a) the lowest state and (b) center state of the spectrum for different $\lambda$, where $n$ is the subscript of the Fibonacci numbers $F_n$. (c) $\alpha_{min}$ versus eigenvalues $E$ with fixed $L=F_{17}=2584$ for $\lambda=0.5$ (black dots) and $\lambda=2$ (blue dots). The two red (green) dashed lines represent the MEs $E_c=\pm 2$ ($\pm 0.5$).}
\end{figure}

\section{II. Global theory of one-frequency cocycle}
 Suppose that  $A$ is an analytic function form the circle $S^1$  to  the group $SL(2,C)$,
 an analytic  cocycle $(\omega, A)$ is a linear skew product:
\begin{eqnarray*}\label{cocycle}
(\omega,A):& S^{1} \times R^2 \to S^{1} \times R^2\\
\nonumber &(\theta,v) \mapsto (\theta+\omega,A(\theta) \cdot v).
\end{eqnarray*}
If $A(\theta)$ admits a holomorphic extension to $|\Im \theta|<\delta$, then for
$|\epsilon|<\delta$ we can define $A_\epsilon(\theta)=A(\theta+i \epsilon)$, and define its Lyapunov exponent by $$\gamma_{\epsilon}(A)=\lim_{n\rightarrow \infty}\frac{1}{n} \int \ln  \|A_n(\theta +i \epsilon)\| d\theta,$$  where $A_n$ is the transfer matrix.
The key observation of Avila's global theory~\cite{A4S} is that  $\epsilon \rightarrow \gamma_{\epsilon}(A)$
is convex and piecewise linear, with right-derivatives satisfying
\begin{equation*}
\lim_{\epsilon \rightarrow 0+}  \frac{1}{2\pi \epsilon} (\gamma_{\epsilon}(A)-\gamma(A))  \in \mathbb{Z}.
\end{equation*}

Note that  in our case,  a sequence $(u_n)_{n \in \mathbb{Z}}$ is a formal solution of the
eigenvalue equation $u_{n+1}+u_{n-1}+v(n)u_n=Eu_n$ if and only if
it satisfied $\begin{pmatrix}
u_{n+1}\\u_n\end{pmatrix}= \begin{pmatrix}
 E-v(n) &  -1\cr
  1 & 0 \end{pmatrix} \begin{pmatrix} u_n\\u_{n-1} \end{pmatrix}$, while the operator can be seen as a cocycle, however, the cocycle is not analytic since the potential is not a smooth function.
The useful observation is that its iterate $T_{\kappa}$ can be seen as   an analytic  cocycle $(\kappa \omega, T_{\kappa}(\cdot))$. Thus by the general theory,  the Lyapunov exponent of the cocycle    $\gamma_{\epsilon}(T_{\kappa}) = \kappa\gamma_{\epsilon}(E)$ is a convex, piecewise linear function, their slopes are integers multiply $2\pi$.

\section{III. Localization length}
\begin{figure}[h]
\hspace*{-0.1cm}
\includegraphics[width=0.95\textwidth]{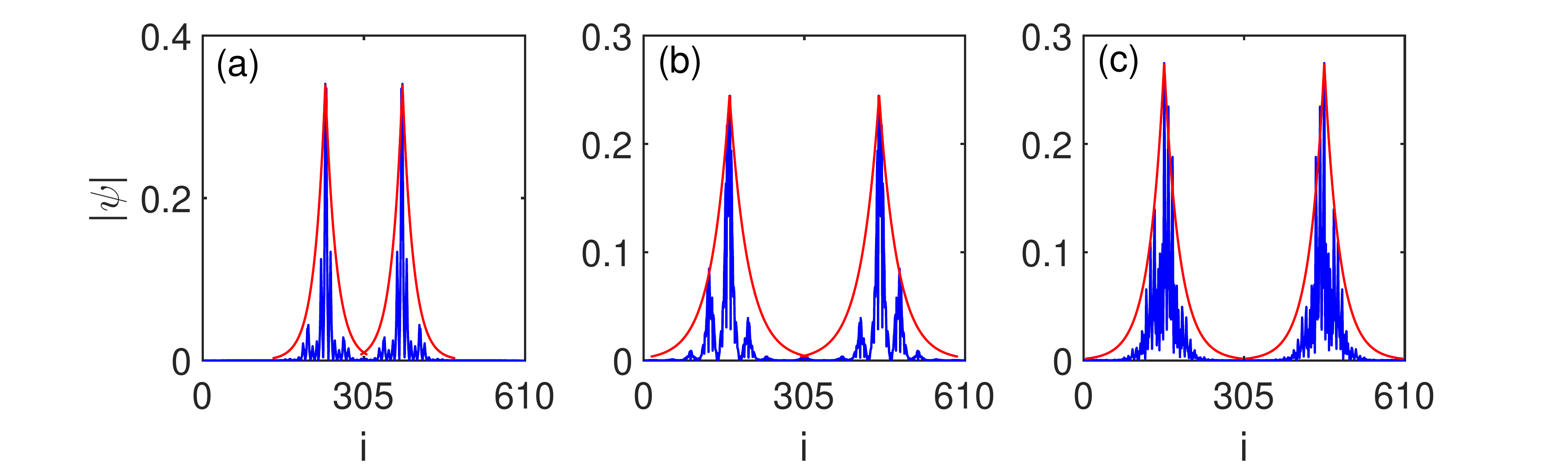}
\caption{\label{length}
Blue lines represent the spatial distributions of $|\psi|$, where $\psi$ is the eigenstate corresponding to (a) $-2.205(9)$ with $\lambda=0.5$ (i.e., corresponding Fig.3(a) of the main text), (b) $-1.764(6)$ with $\lambda=0.6$, and (c) $-1.071(9)$ with $\lambda=1$. Here we fix $L=F_{15}=610$ and $\kappa=2$. The red lines are $|\psi|_{max} exp(-|i-i_0|/\xi)$.}
\end{figure}

 In the main text, our analytical solutions provide the exact results not only for the MEs, but also for the localization lengths of all localized states, i.e., $\xi(E)=\frac{\kappa}{\ln \lambda a_{\kappa}}$, as showed in Eq.(7) of the main text. In this section, we numerically verify this theoretical expressions, as showed in Fig. \ref{length}. The red lines represent $|\psi|_{max} exp(-|i-i_0|/\xi)$, where $|\psi|_{max}$ are the maximum values of $|\psi|$ in the two peaks, $i_0$ are the corresponding lattice sites and $\xi$ are the localization length satisfying $\xi(E)=\frac{\kappa}{\ln |\lambda a_{\kappa}|}$.
From these figures, we see that the analytical expressions of the localization length can well describe the localization features of the corresponding eigenstates.

\section{IV. Experimental Realization}

\begin{figure}
\includegraphics[width=0.55\textwidth]{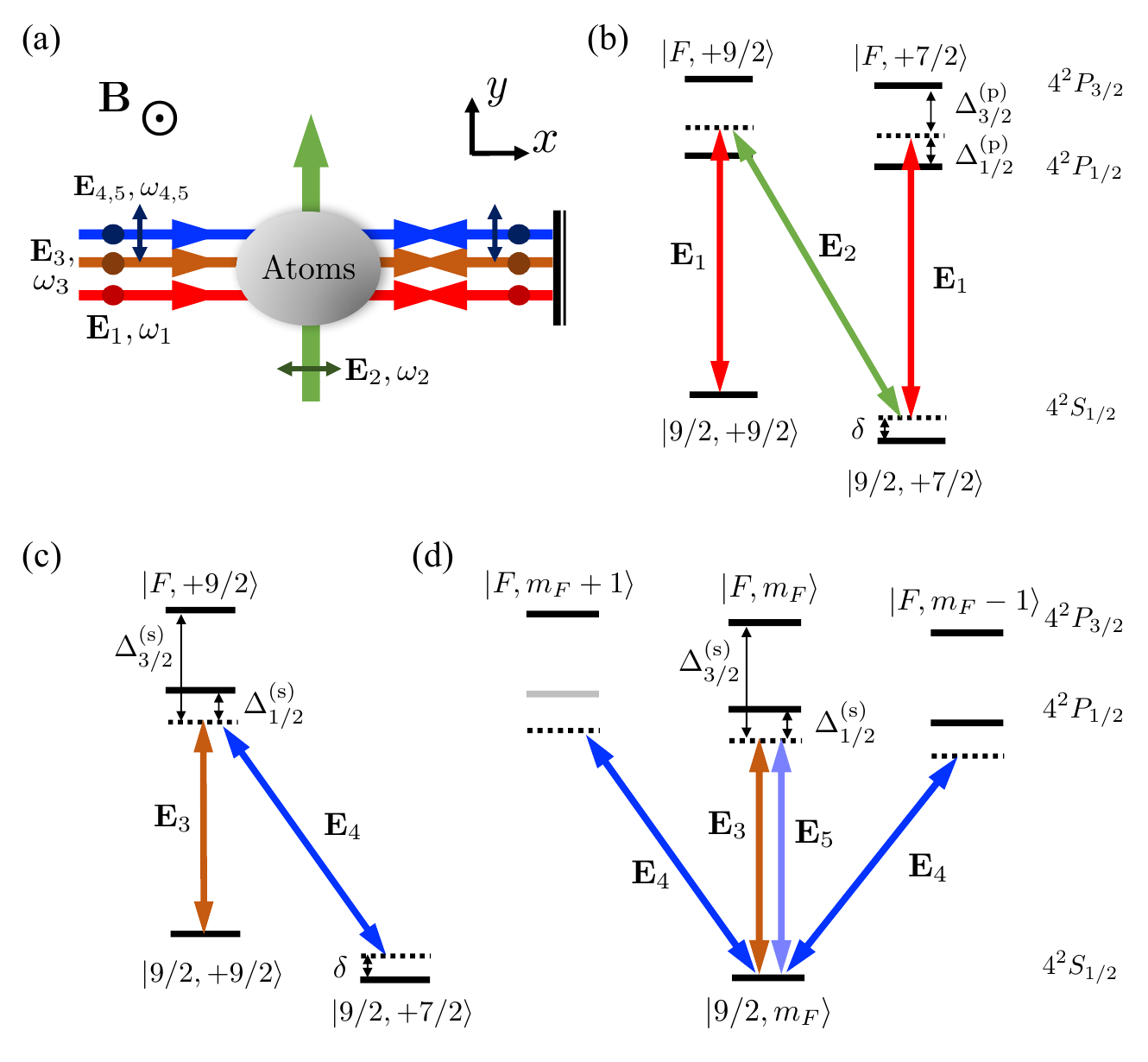}
\caption{Experimental setup and light couplings for $^{40}$K atoms.
(a) Schematic of experimental setup. A standing wave ${\bf E}_1$ of frequency $\omega_1$ with $z$ polarization and
a plane wave ${\bf E}_2$ of frequency $\omega_2$ with $x$ polarization generate a Raman coupling potential [see (b)], which plays the role of the prime lattice.
Two standing waves ${\bf E}_3$ of frequency $\omega_3$ with $z$ polarization and ${\bf E}_4$ of frequency $\omega_4$ with $y$ polarization
generate another Raman coupling potential [see (c)], which provides the secondary incommensurate lattice.
Another standing-wave beam ${\bf E}_5$ of frequency $\omega_5\sim\omega_4$ with $z$ polarization is applied to produce a lattice potential which eliminates
the incommensurate potential for spin-up atoms [see (d)]. Here we set $\omega_1-\omega_2=\omega_3-\omega_4$.
(b) Optical transitions for generating the prime lattice.
(c-d) Optical transitions for the secondary incommensurate lattice.
}\label{figS1}
\end{figure}

In this section, we illustrate how to to realize the lattice model (8) in the main text.
Our basic idea is to use well-tuned Raman couplings to generate both the primary and secondary lattice potentials.
We shall first realize the Hamiltonian
\begin{align}\label{Hlattice_S}
H=\frac{k_x^2}{2m}\otimes\id+\frac{V_p}{2}\cos(2k_px)\sigma_x-M_0\sigma_z+\frac{V_s}{4}\cos(2k_sx)(-\id+\sigma_x),
\end{align}
which can be transformed into the form (8) under the spin rotation
\begin{align}\label{rotation_S}
\sigma_x\to\sigma_z,\quad\sigma_z\to-\sigma_x.
\end{align}
Our proposed experimental setup is sketched in Fig.~\ref{figS1}(a).
In the following we shall take $^{40}$K atoms as an example while all our results are applicable to other alkali atoms. For $^{40}$K, the spin-$1/2$ system can be constructed
by $|\!\uparrow\rangle=|F=9/2, m_F=+9/2\rangle$ and $|\!\downarrow\rangle=|9/2, +7/2\rangle$.
The lattice and Raman coupling potentials are contributed from both the $D_2$ ($4{^{2}S}_{1/2}\to4{^{2}P}_{3/2}$) and $D_1$ ($4^{2}S_{1/2}\to4^{2}P_{1/2}$) lines [Fig.~\ref{figS1}(b-d)].

\subsubsection{A. The primary lattice}

The primary lattice is generated by a Raman coupling via a standing wave field ${\bf E}_{1}=2E_{1}\hat{z}e^{\ui(\phi_1+\phi_{1e}/2)}\cos(k_1 x-\phi_{1e}/2)$ of frequency $\omega_1$
and a plane wave ${\bf E}_2=\hat{x}E_2 e^{\ui (k_1y+\phi_2)}$ of frequency $\omega_2$,
where $\phi_{1,2}$ denote the initial phases, and $\phi_{1e}$  is the phase acquired by ${\bf E}_1$ for an additional optical path
back to the atom cloud.
As shown in Fig.~\ref{figS1}(b), the standing-wave field ${\bf E}_{1}$ creates a lattice ${\cal V}_1(x)$,
which is given by
\begin{align}
{\cal V}_{1\sigma}(x)=\sum_{F}\frac{\left|\Omega^{(3/2)}_{\sigma F,1z}\right|^2}{\Delta^{\rm (p)}_{3/2}}+\sum_{F}\frac{\left|\Omega^{(1/2)}_{\sigma F,1z}\right|^2}{\Delta^{\rm (p)}_{1/2}},
\end{align}
where $\Omega_{\sigma F,1z}^{(J)}=\langle\sigma|er|F,m_{F\sigma},J\rangle\hat{z}\cdot{\bf E}_{1}$  
($J=1/2,3/2$). From the dipole matrix elements of
$^{40}$K~\cite{Potassium}, we obtain
\begin{align}
&{\cal V}_{1}(z)=V_1\cos^2(k_1 z-\phi'_1/2),\nonumber\\
&V_{\rm 1}=\frac{4t_{1/2}^2}{3}\left(\frac{2}{|\Delta^{\rm (p)}_{3/2}|}-\frac{1}{|\Delta^{\rm (p)}_{1/2}|}\right)E^2_{1},
\end{align}
with the transition matrix elements $t_{1/2}\equiv\langle J=1/2||e{\bf r}||J'=1/2\rangle$, $t_{3/2}\equiv\langle J=1/2||e{\bf r}||J'=3/2\rangle$ and $t_{3/2}\approx\sqrt{2}t_{1/2}$.
The Raman coupling potential via ${\bf E}_{1,2}$ is
 \begin{align}
{\cal M}_{12}(x)=\sum_{F}\frac{\Omega^{(3/2)*}_{\uparrow F,1z}\Omega^{(3/2)}_{\downarrow F,2+}}{\Delta^{\rm (p)}_{3/2}}
 +\sum_{F}\frac{\Omega^{(1/2)*}_{\uparrow  F,1z}\Omega^{(1/2)}_{\downarrow F,2+}}{\Delta^{\rm (p)}_{1/2}},
\end{align}
with $\Omega_{\sigma F,2+}^{(J)}=\langle\sigma|er|F,m_{F\sigma}+1,J\rangle\hat{e}_+\cdot{\bf E}_{2}$,
and takes the form ${\cal M}_{12}(x)=M_{12}\cos(k_1 x-\phi_{1e}/2)e^{\ui (k_1y+\phi_2-\phi_1-\phi_{1e}/2-\pi)}$, where
 \begin{align}
 M_{12}=\frac{2t_{1/2}^2}{9}\left(\frac{1}{|\Delta^{\rm (p)}_{1/2}|}+\frac{1}{|\Delta^{\rm (p)}_{3/2}|}\right)E_1E_2.
 \end{align}

We assume that the wavelength of ${\bf E}_{1}$ is $\lambda_1=769$nm, which satisfy $\Delta^{\rm (p)}_{3/2}=-2\Delta^{\rm (p)}_{1/2}=2\Delta_{\rm FS}/3$
where $\Delta_{\rm FS}$ denotes the fine structure splitting.
We then have $V_{\rm 1}=0$, and $M_{12}=t_{1/2}^2E_1E_2/\Delta_{\rm FS}$. We further set $\phi_2-\phi_1-\phi_{1e}/2=\pi+2n\pi$ ($n=0,1,2,\cdots$).
Under the spin rotation (\ref{rotation_S}), we have the primary lattice ${\cal V}_p(x)=\frac{V_p}{2}\cos(2k_p x-\phi_{1e}/2)$,
with
 \begin{align}
V_p=2t_{1/2}^2E_1E_2/\Delta_{\rm FS}, \quad  k_p=k_1/2.
 \end{align}
 Moreover, we set a non-zero two-photon detuning $\delta=\Delta\omega_z-\omega_1+\omega_2$, where $\Delta\omega_z$ denotes the energy difference between the two spin states, leads to an effective Raman coupling $M_0\sigma_x$ with $M_0=-\delta/2$.

\subsubsection{B. The incommensurate lattice}

We apply three standing wave fields together to generate an incommensurate lattice only for spin-down atoms,
which are ${\bf E}_{3}=2E_{3}\hat{z}e^{\ui(\phi_3+\phi_{3e}/2)}\cos(k_3 x-\phi_{3e}/2)$
of frequency $\omega_3$,
${\bf E}_{4}=2E_{4}\hat{y}e^{\ui(\phi_4+\phi_{4e}/2)}\cos(k_3 x-\phi_{4e}/2)$ of frequency  $\omega_4$,
${\bf E}_{5}=2E_{5}\hat{z}e^{\ui(\phi_5+\phi_{5e}/2)}\cos(k_3 x-\phi_{5e}/2)$ of frequency  $\omega_5\sim\omega_4$,
where $\Delta\omega_z-\omega_3+\omega_4=\delta$.
These standing waves create three spin-independent lattices ${\cal V}_{j}(x)=V_j\cos^2(k_3 x-\phi_{je}/2)$ ($j=3,4,5$) [see Fig.~\ref{figS1}(d)] with
\begin{align}
V_j=\frac{4t_{1/2}^2}{3}\left(\frac{2}{|\Delta^{\rm (s)}_{3/2}|}+\frac{1}{|\Delta^{\rm (s)}_{1/2}|}\right)E^2_{j}.
\end{align}
If $\phi_{3e}-\phi_{4e}=(2n+1)\pi$ ($n$ is an integer) and $E_3=E_4$, we have ${\cal V}_{3}(x)+{\cal V}_{4}(x)={\rm const.}$.
The Raman coupling potential via ${\bf E}_{3,4}$ is ${\cal M}_{34}(x)=-M_{34}\sin(2k_3 x-\phi_{3e})$ (assuming $\phi_4-\phi_3=(n+1+2m)\pi$ with $m$ being an integer),
where
\begin{align}
M_{34}=\frac{2t_{1/2}^2}{9}\left(\frac{1}{|\Delta^{\rm (s)}_{1/2}|}-\frac{1}{|\Delta^{\rm (s)}_{3/2}|}\right)E_3E_4.
\end{align}

We further assume $\phi_{5e}-\phi_{3e}=\pi/2+2n\pi$ and $V_{\rm 5}=2M_{34}$; the latter can be achieved by tuning $\Delta^{\rm (s)}_{1/2,3/2}$ and $E_{5}$.
Hence, under the rotation (\ref{rotation_S}), we have the secondary incommensurate  lattice ${\cal V}_s(x)=\frac{V_s}{2}\sin(2k_s x-\phi_{3e}+\pi)$ felt only by the spin-down state,
with
 \begin{align}
V_s=\frac{2t_{1/2}^2}{9}\left(\frac{1}{|\Delta^{\rm (s)}_{1/2}|}-\frac{1}{|\Delta^{\rm (s)}_{3/2}|}\right)E_3^2, \quad  k_s=k_3.
 \end{align}

 \begin{figure}
\includegraphics[width=0.4\textwidth]{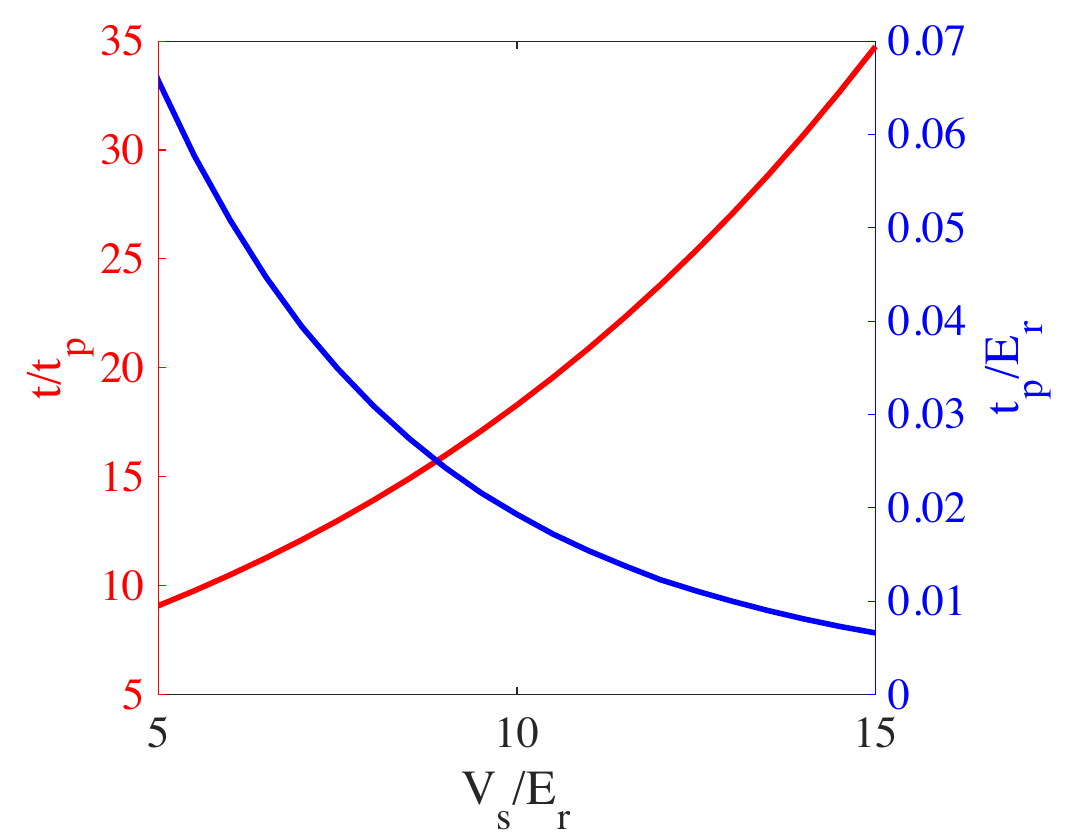}
\caption{Spin-conserved and -flipped hoppings versus the lattice depth $V_s$. Here we set $M_0=1.5E_r$.
}\label{figS2}
\end{figure}

\subsubsection{C. Tight-binding model}

In the tight-binding limit and only considering $s$-bands, the spin-conserved hopping induced by the primary lattice is
$t_p=-\int dx\phi_{s}(x)\left[\frac{k_x^2}{2m}+{\cal V}_{p}(x)\right]\phi_{s}(x-a)$,
where $a=\pi/k_1$ is the lattice period of the primary lattice and $\phi_{s}(x)$ denotes the Wannier function.
When $t_p$ is negligible,
the Hamiltonian (8) can take the form of Eq.~(2) with
the spin-flipped hopping playing the role of nearest-site tunneling, i.e.
\begin{align}
t=M_0\int dx\phi_{s}(x)\phi_{s}(x-a/2),
\end{align}
and
\begin{align}
\begin{split}
\lambda_j&=\frac{1}{2}\int dxV_{s}(x)|\phi_{s}(x-ja)|^2=\lambda\cos(2\pi\omega j-\phi_{3e}-\pi/2),\\
\lambda&\equiv\frac{V_s}{4}\int dx\cos(2k_3x)|\phi_{s}(x)|^2,
\end{split}
\end{align}
with $\omega=\{2k_3/k_1\}$. Here $\{\cdot\}$ denotes the fractional part.
We define the recoil energy $E_r\equiv k_p^2/(2m)=k_1^2/(8m)$, and calculate both spin-conserved and -flipped hoppings as a function of $V_s$.
The results are shown in Fig.~\ref{figS2}. We find that when setting $M_0=1.5E_r$, $V_s=10E_r$ is deep enough to meet our need.


\end{document}